\documentstyle[aps,prl,preprint,floats,epsfig]{revtex}
\newcommand{\mydef}[2]{\def #1{\ifmmode{\rm #2}
          \else ${\rm #2}$\fi}}
\newcommand{\Res}[3]{\ifmmode{#1\:_{\footnotesize -#2}^{\footnotesize +#3}\mbox{\rule{0cm}{2.5ex}}} \else
$#1\:_{\footnotesize -#2}^{\footnotesize +#3}$\rule{0cm}{2.5ex}\fi}
\mydef{\Pgpp}{\pi^+}
\mydef{\PDm}{D^-}
\newcommand{\etal}{\mbox{$et$ $al.$}}
\def\aDz{\ifmmode{\mathrm{\overline{D}}^{\: 0}}\else ${\mathrm{\overline{D}}^{\: 0}}$\fi}
\def\Dz{\ifmmode{\mathrm{D}^{0}}\else ${\mathrm{D}^{0}}$\fi}
\def\gDz{\Dz}
\def\aDstz{\ifmmode{\mathrm{\overline{D}}^{\:\ast 0}}\else ${\mathrm{\overline{D}}^{\:\ast 0}}$\fi}
\def\Dstz{\ifmmode{\mathrm{D}^{\ast 0}}\else ${\mathrm{D}^{\ast 0}}$\fi}
\def\Bz{\ifmmode{{\mathrm{B}^0}}\else ${\mathrm{B}^0}$\fi}
\def\aBz{\ifmmode{\mathrm{\overline{B}}^0}\else ${\mathrm{\overline{B}}^0}$\fi}
\def\gpz{\ifmmode{\mathrm{\pi^0}}\else ${\mathrm{\pi^0}}$\fi}
\def\lamplace{\rule{0cm}{2.5ex}}
\mydef{\degree}{^\circ}
\mydef\BR{{\cal B}}
\def\myfigwidth{155mm}
\newcommand{\up}[1]{\raisebox{1.5ex}[-1.5ex]{#1}}
\mydef{\myarrow}{{\rightarrow}}
\mydef{\kreuz}{\times}
\mydef{\yfours}{\Upsilon(4S)}
\mydef{\Ebeam}{E_{beam}}
\mydef{\Dtokpi}{\gDz\myarrow K^- \pi^+}
\mydef{\Dtokpipi}{\gDz\myarrow K^- \pi^+ \gpz}
\mydef{\Dtokpipipi}{\gDz\myarrow K^- \pi^+\pi^+\pi^-}
\mydef{\Dtokthreepi}{\gDz\myarrow K^-(3\pi)^+}
\mydef{\Kpi}{K^- \pi^+}
\mydef{\Kpipi}{K^- \pi^+ \gpz}
\mydef{\Kpipipi}{K^- \pi^+\pi^+\pi^-}
\mydef{\mbc}{M_{B}}
\mydef{\mbcc}{\mbc}
\mydef{\deltae}{\Delta E}
\mydef{\FD}{{\cal F}_D}
\mydef{\ctst}{\cos\theta_{B Hel.}}
\mydef{\Btodpi}{\aBz\myarrow\Dz\gpz}
\mydef{\Btodstpi}{\aBz\myarrow\Dstz\gpz}
\mydef{\BBar}{B\overline{B}}
\mydef{\Dsttodpi}{\Dstz\myarrow\Dz\gpz}
\mydef{\Dsttodg}{\Dstz\myarrow\Dz\gamma}
\mydef{\Navfit}{\overline{N_{fit}}}
\mydef{\Nexp}{N_{{\footnotesize Expect.}}}
\mydef{\RMSfit}{{\small RMS}}
\mydef{\Significance}{\frac{N}{\sigma_N}}
\mydef{\Nsig}{N_{Sig}}
\mydef{\Nbb}{N_{B\overline{B}}}
\mydef{\Ncont}{N_{Cont}}
\mydef{\Like}{{\cal L}}
\mydef{\DDstz}{D^{(*)0}}
\mydef{\Btoddstpi}{\aBz\myarrow\DDstz\gpz}
\mydef{\Efficiency}{\varepsilon}
\mydef{\rhm}{\rho^-}
\mydef{\Btoddstrho}{B^-\myarrow D^{(*)0}\rho^-}
\mydef{\Btodrho}{B^-\myarrow D^{0}\rho^-}
\mydef{\Btodstrho}{B^-\myarrow D^{*0}\rho^-}

\def\ppicut{1.8 GeV}
\def\pdcut{1.65 GeV}
\def\pdstcut{1.8 GeV}

\def\dsignalregion{$-0.05$ $<$ \deltae\ $<$ $0.05$ GeV, $5.275$ $<$ \mbc\
$<$ $5.285$ GeV}

\mydef{\BRdpi}{{2.74}_{\footnotesize -0.32}^{\footnotesize +0.36}}
\mydef{\BRdstpi}{{2.20}_{\footnotesize -0.52}^{\footnotesize +0.59}}
\def\syserrBdpi{0.55}
\def\syserrBdstpi{0.79}

\def\Bdpisig{12.1}
\def\Bdstpisig{5.9}
\def\cosdi{0.89$\pm$0.08}
\def\cosdistar{0.89$\pm$0.08}

\def\amplratiodpi{0.70$\pm$0.11}
\def\amplratiodstpi{0.74$\pm$0.08}

\def\Bdpisigsyst{9.4}
\def\Bdstpisigsyst{4.2}

\def\efa{37.1} 
\def\efb{13.5} 
\def\efc{19.0} 
\def\efd{15.3} 
\def\efe{5.5} 
\def\eff{8.1} 
\def\efg{11.4} 
\begin{document}
\preprint{\tighten\vbox{\hbox{\hfil CLNS 01/1755}
                        \hbox{\hfil CLEO 01-18}}}

\title{\large Observation of \Btodpi\ and \Btodstpi}
\author{(CLEO Collaboration)}
\date{January 11, 2002}
\maketitle
\tighten

\begin{abstract}
We have studied the color-suppressed hadronic decays of neutral B mesons into
the final states D$^{(*)0}$\gpz.
Using 9.67 million \BBar\ pairs collected with the CLEO detector, we
observe the decays \Btodpi\ and \Btodstpi\ with the
branching fractions  \BR(\Btodpi) = (\BRdpi$\pm$\syserrBdpi)\kreuz
10$^{-4}$ and \BR(\Btodstpi) = (\BRdstpi$\pm$\syserrBdstpi)\kreuz 10$^{-4}$.
The first error is statistical and the second systematic. The
statistical significance of the \Dz\gpz\ signal is \Bdpisig$\sigma$
(\Bdstpisig$\sigma$ for \Dstz\gpz). Utilizing the \Btoddstpi\ branching
fractions we determine the strong phases $\delta_{I,D(*)}$ between isospin 1/2 and 3/2
amplitudes in the ${\rm D\pi}$ and ${\rm D^*\pi}$ final states 
to be $\cos \delta_{I,D}$=\cosdi\ 
and $\cos \delta_{I,D*}$=\cosdistar, respectively.
\end{abstract}
\begin{figure}[b]
\vspace*{-0.5cm}
\centerline{\epsfig{file=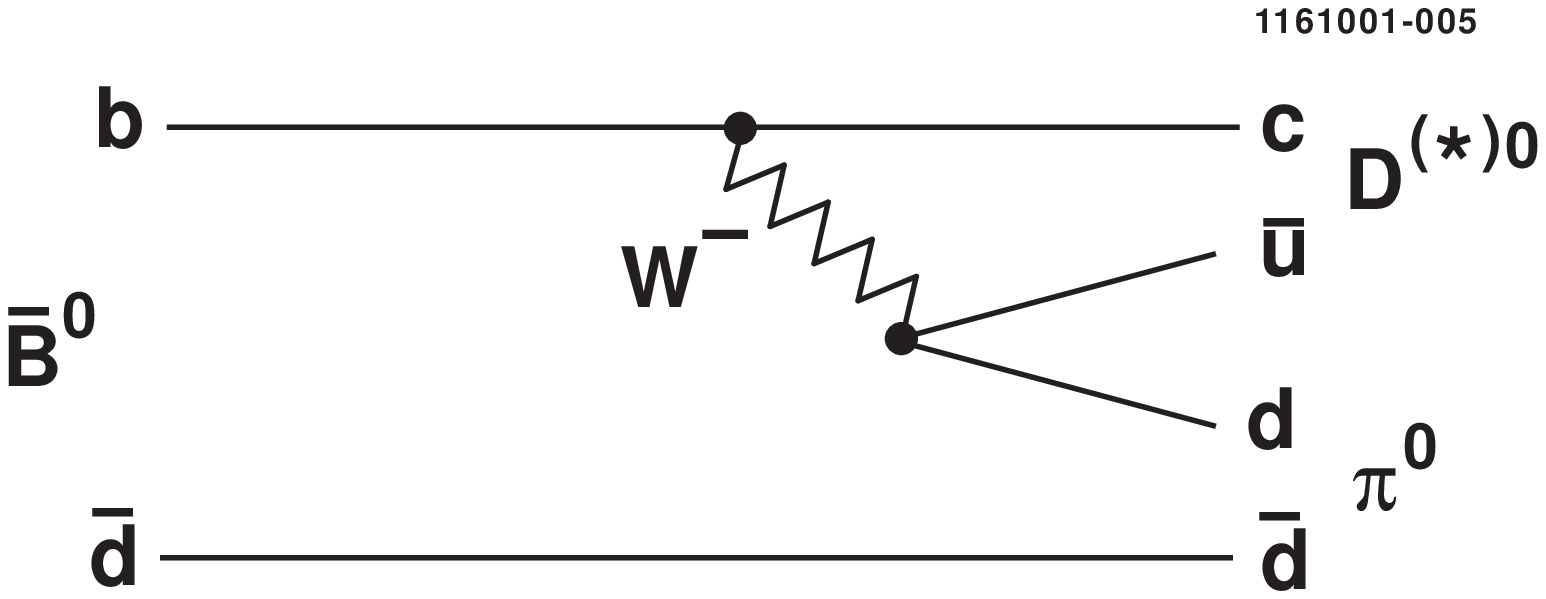,width=10cm}}
\vspace*{-0.3cm}
\begin{center}
\parbox{145mm}{\caption[ ]{\label{feynman.eps}Diagram for the
color-suppressed decay \Btoddstpi.}}
\end{center}
\end{figure}

\pacs{13.25.Hw}
\begin{center}
{
T.~E.~Coan,$^{1}$ Y.~S.~Gao,$^{1}$ F.~Liu,$^{1}$
Y.~Maravin,$^{1}$ I.~Narsky,$^{1}$ R.~Stroynowski,$^{1}$
J.~Ye,$^{1}$
M.~Artuso,$^{2}$ C.~Boulahouache,$^{2}$ K.~Bukin,$^{2}$
% ...
}
\end{center}
{\small
\begin{center}
$^{1}${Southern Methodist University, Dallas, Texas 75275}\\
$^{2}${Syracuse University, Syracuse, New York 13244}\\
% ...
$^{4}${Vanderbilt University, Nashville, Tennessee 37235}\\
% ...
$^{13}${University of Illinois, Urbana-Champaign, Illinois 61801}\\
% ...
\end{center}}
\clearpage
The decay \Btoddstpi\ proceeds predominantly through the internal
spectator diagram shown in Fig.~\ref{feynman.eps}. 
This diagram is color-suppressed,  
since the color of the quark-pair originating from the W 
decay must match the color of the other quark pair.
The rate of \Btoddstpi\ relative to ${\rm B^-\myarrow\DDstz\pi^-}$
is suppressed, crudely, by one factor of (1/3)$^2$, and an additional 
$(1/\sqrt{2})^2$ due to the projection of the ${\rm d\overline{d}}$ 
state onto the pion wave function \cite{browder_honscheid}. 
This gives a total suppression of 1/18 compared to the color-favored decay
modes.
Detailed theoretical calculations \cite{heavy_flavors}
predict an even larger suppression of about a factor 1/50.

So far the only established color-suppressed decays are two-body 
B decays into Charmonium plus neutral hadrons. 
A measurement of \Btoddstpi\ is therefore a benchmark test for theoretical
models of hadronic B decays\cite{browder_honscheid,heavy_flavors}.
An investigation of color-suppressed decays into a D meson and light neutral 
mesons other than a \gpz\ is currently underway and will be addressed in a future publication.

The observation of \Btoddstpi\ completes the measurement of ${\rm
D^{(*)}\pi}$ final states and allows us to extract the strong phase difference
between isospin 1/2 and 3/2 amplitudes \cite{heavy_flavors,rosner99}. 
 
In this Letter we present the observation of \Btoddstpi, 
superseding the limits from our previous publication 
\cite{bdpi0_cleo2}. Our new analysis has significantly 
increased statistics and is based on a data sample with improved
calibration and track reconstruction. 
The background shapes and the signal separation power have also been significantly
improved. Charge conjugates are implied throughout the paper.
Our analysis uses e$^+$e$^-$ annihilation data recorded with the CLEO
detector at the Cornell Electron Storage Ring.
The integrated luminosity of our data sample is 9.15 fb$^{-1}$ for data collected 
on the \yfours\ (on-resonance), corresponding to 9.67 million \BBar\ pairs,
and 4.35 fb$^{-1}$ collected 60 MeV below the \BBar\ threshold
(off-resonance), which is used for background studies. 

CLEO is a general purpose solenoidal magnet detector. 
Data were recorded with two detector configurations, CLEO II and CLEO II.V 
\cite{cleo2_detector,cleo2.5_detector}.
Cylindrical drift chambers in a 1.5T solenoidal magnetic field measure momentum
and specific ionization (dE/dx) of charged particles. Photons are
detected using a CsI(Tl) crystal electromagnetic calorimeter,
consisting of a barrel-shaped central part of 6144 crystals and
1656 crystals in the forward regions of the detector (endcaps).
In the II.V configuration the innermost chamber was replaced by a
three-layer, double-sided silicon microvertex detector, and the
main drift chamber gas was changed from argon-ethane to a 
helium-propane mixture. As a result of these modifications, the CLEO II.V part
of the data (2/3 of the total sample) has improved momentum resolution
and particle identification.

B mesons were reconstructed by selecting high-momentum 
D$^{(*)0}$ and \gpz\ mesons. 
Track quality requirements are imposed on charged tracks and the
purity of pion and kaons is improved by using dE/dx information
whenever available. The \gpz\ candidates are reconstructed from isolated
electromagnetic clusters of at least 30 MeV in the barrel region and 50
MeV in the endcaps. The mass resolution is 8 MeV in the barrel region and 
10 MeV in the endcaps.
We require that the candidate's
mass is within $2.5$ standard deviations ($\sigma$) of the nominal
\gpz\ mass. Prompt \gpz's from B decays are required 
to have a momentum larger than \ppicut.

\Dz\ mesons are selected in the decay modes \Dtokpi, \Dtokpipi\ and
\Dtokpipipi. The invariant mass of the D daughter particles is
required to be within $2.5$ $\sigma$ of the
known \Dz\ mass. The mass resolution depends on the decay mode and is between 6 and 12 MeV. The momentum of the \Dz\ is required to be larger than
\pdcut. In the \Dtokpipi\ mode we suppress combinatorial 
background by using only certain regions of the Dalitz plane.

\Dstz\ mesons are selected in the decay modes \Dsttodpi\ and \Dsttodg.
To reduce the combinatorial background in the 
decay mode \Dsttodg\ we require that the \Dz\ decays into 
${\rm K^- \pi^+}$.
We require that the mass difference m$_\Dstz$-m$_\Dz$ is within
$2.5\sigma$ of the known value and that the
momentum of the \Dstz\ is larger than \pdstcut. 
The kinematic resolution of \gpz\ and D$^{(*)0}$ candidates
is improved by a mass-constrained kinematic fit.

B decay candidates are selected from \gpz\ and D$^{(*)0}$ pairings
that have no electromagnetic clusters in common. 
B candidates are identified using a beam-constrained mass 
${\rm \mbc=\sqrt{\Ebeam^2-p_B^2}}$, where \Ebeam\ is the beam energy and
${\rm p_B}$ is the B candidate momentum, and an energy difference
${\rm \deltae = E_{D} + E_\gpz - \Ebeam}$, where ${\rm E_{D}}$ and
${\rm E_\gpz}$ are the energies of the D$^{(*)0}$ and \gpz. The
resolution in \mbc\ depends on the D decay mode and is 
between 3.5 and 4 MeV. The resolution is
dominated by the beam energy spread and the \gpz\ energy
resolution. The resolution in \deltae\ is between 35 and 40 MeV.
The energy resolution is slightly asymmetric due to the energy loss
out of the back of the CsI crystals. The mass-constrained kinematic
fit to the pion 4-momentum compensates for most of this effect.
We accept B candidates with \mbc\ above 5.24 GeV and $|\deltae|$ $<$
300 MeV.
To better suppress background from ${\rm e^+e^-\rightarrow q\bar{q}}$ events 
(continuum background), several event shape variables
are combined into a Fisher discriminant \FD \cite{cleo2_pipikk}. 
For \BBar\ events (continuum background), the \FD\ distribution 
is almost a Gaussian and has its maximum 
at 0.42 (0.57). The
standard deviation is 0.11 (0.12). The separation between the \BBar\ and continuum distributions is 1.3 $\sigma$.
We reject clear continuum events by requiring \FD $<$ 1. 
For each candidate we calculate the sphericity vectors\cite{sphericity} of the B
daughter particles and of the rest of the event. 
We require the cosine of the angle between these two vectors to be within -0.8 and 0.8.
The distribution of this angle is strongly peaked at $\pm$1 for
continuum background and is nearly flat for \BBar\ events.

The number of signal events in the sample is obtained from 
unbinned, extended maximum likelihood fits. The free parameters 
of the fits are
the number of signal events, background from B decays  (\BBar) and 
from continuum e$^+$e$^-$ annihilation (continuum).
Four variables are used as input to the maximum likelihood fit:
the beam-constrained mass \mbc, the energy difference \deltae,
the Fisher Discriminant \FD, and the cosine of the decay angle of
the B \ctst, defined as the   
angle between the D$^{(*)}$ momentum and the B flight direction 
calculated in the B rest frame.

In each of the fits, the likelihood of the B candidate is the sum of
probabilities for the signal and two background hypotheses with
relative weights maximizing the likelihood.
The probability of a particular hypothesis is the product of
probability density functions (PDFs) for each of the input variables.
The PDFs for \mbc\ are represented by a bifurcated Gaussian\cite{bifurcated} for signal,
an empirical shape, $\mbc\:\sqrt{(1-x^2)}\:
\exp(-E_{fact}(1-x^2))$, with $x=\mbc/{\rm E_{beam}}$, for continuum and a Gaussian on top of the empirical 
background
shape\footnote{The parameters of the empirical shapes for \BBar\ and Continuum 
background are different.} 
for \BBar; the PDFs for \deltae\ are the sum of two Gaussians with
a common mean (signal), 1st-order polynomial (continuum) and a
sum of two Gaussians plus a 1st-order polynomial (\BBar);
the PDFs for \FD\ are the sum of two Gaussians with
a common mean; and the PDFs for \ctst\ are 2nd-order polynomials. 
The PDF parameters are determined from off-resonance CLEO data
(continuum) and from high-statistics Monte-Carlo (MC) samples 
(signal and \BBar).

Monte Carlo experiments are generated to test the fitting procedure
and to obtain the relation between fit yield and signal branching
fractions. The experiments are repeated several hundred times with different
Monte Carlo test samples randomly selected from high-statistics MC
samples. 

We summarize the results of the fits to CLEO data in Table \ref{fityield.tab}.
We give results for all B decay modes,
corresponding D decay modes and the combination of all D decay modes.
We combine the results for different D decay modes
by adding the log likelihood as a function of the branching fraction.
Branching fractions for each mode are obtained via
\[{\rm \BR(\Btoddstpi)\:\:=\:\:\frac{{\rm Yield_{fit}}}{\Efficiency \:\kreuz\:
\BR(D^{(*)}) \:\kreuz\:N(\Bz,\aBz)}}\]
The number of \Bz\ plus \aBz, ${\rm N(\Bz,\aBz)}$=9.67 M $\pm$ 0.10 M, 
is derived
assuming equal branching fractions for charged and neutral B meson
decays \cite{fplusminuserror}. 
The uncertainty in the branching fractions   
of the \yfours\ is taken into account in the systematics.
The significances of the observed signals in the seven fits is determined from
the change in -2$\log \Like$ when refit with the signal yield
constrained to zero: $significance \:=\: \sqrt{2(\log\Like-
\log\Like_{\Nsig=0})}$.
We obtain a total significance of \Bdpisig$\sigma$
for \Btodpi\ and \Bdstpisig$\sigma$ for \Btodstpi.
Varying the PDF shapes within the systematic errors to obtain the
lowest signal yield, the statistical
significance is reduced to \Bdpisigsyst$\sigma$
(\Btodpi) and \Bdstpisigsyst$\sigma$ (\Btodstpi).
We obtain branching fractions of 
$\BR(\Btodpi)$ = $(\BRdpi\pm\syserrBdpi)\kreuz 10^{-4}$ and
$\BR(\Btodstpi)$ = $(\BRdstpi\pm\syserrBdstpi)\kreuz 10^{-4}$.
The first error is statistical and the second error systematic.
Our result for \Btodpi\ is higher than the previous CLEO upper limit 
\cite{bdpi0_cleo2}. We ascribe this disagreement, which is of the order of
$3.1$ $\sigma$, partly to a statistical fluctuation and 
partly to the description of the \deltae-background in the old CLEO
publication.

\begin{figure}[htb]
\centerline{\epsfig{file=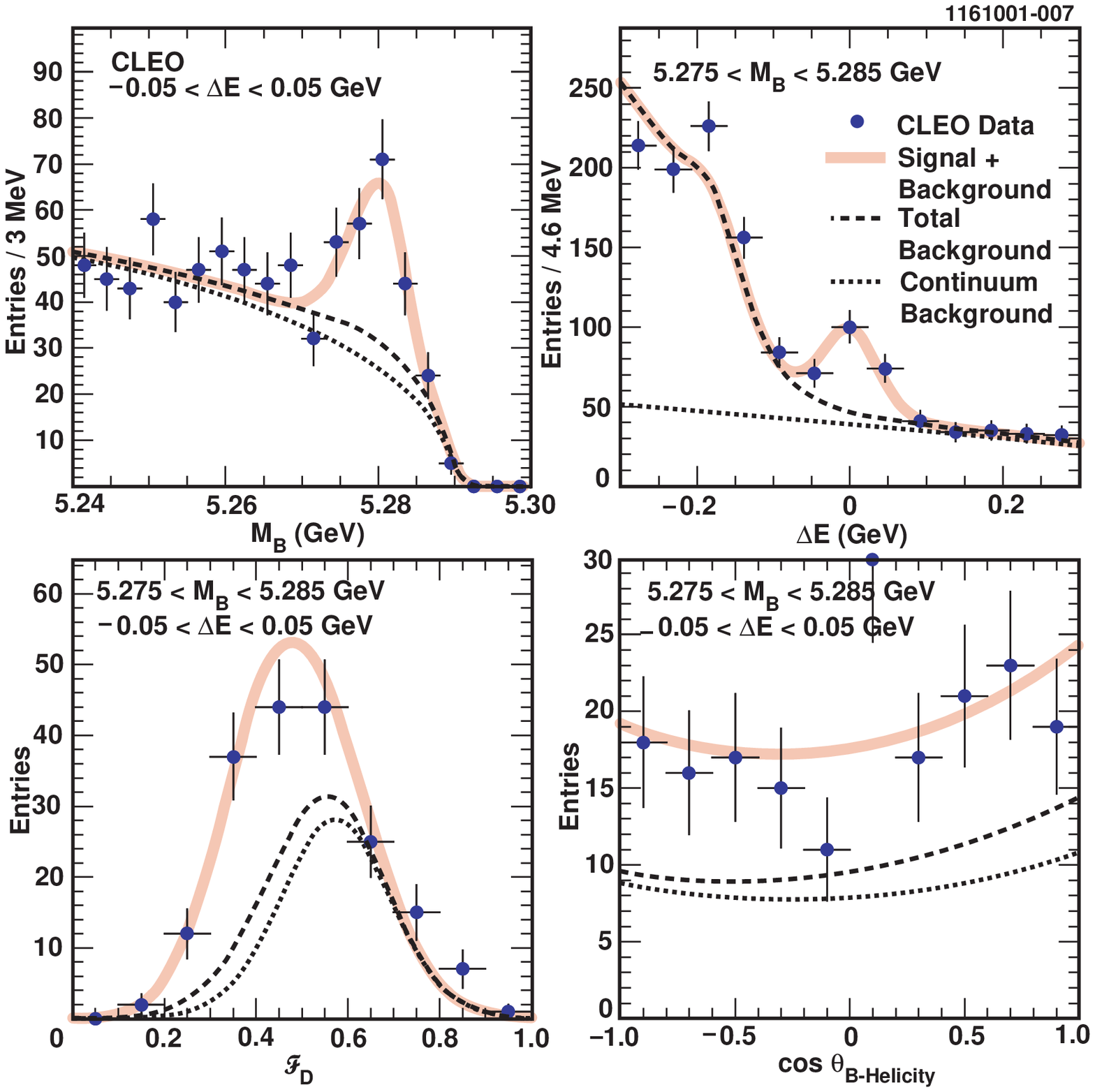,width=\myfigwidth}}
\begin{center}
\parbox{145mm}{\caption[ ]{\label{signalregion_overlay_bdpi0.eps}\it Distribution of fit
input variables for \Btodpi. The results of the unbinned, extended
maximum likelihood fit are shown as the 
full line. The dotted line represents the fitted 
continuum and the dashed line is the fit result for the sum of \BBar\ 
and continuum background. 
To enhance the signal for display purposes, the fit results are projected 
into the \mbc-\deltae\ signal region \dsignalregion.}}
\end{center}
\end{figure}

\begin{figure}[htb]
\centerline{\epsfig{file=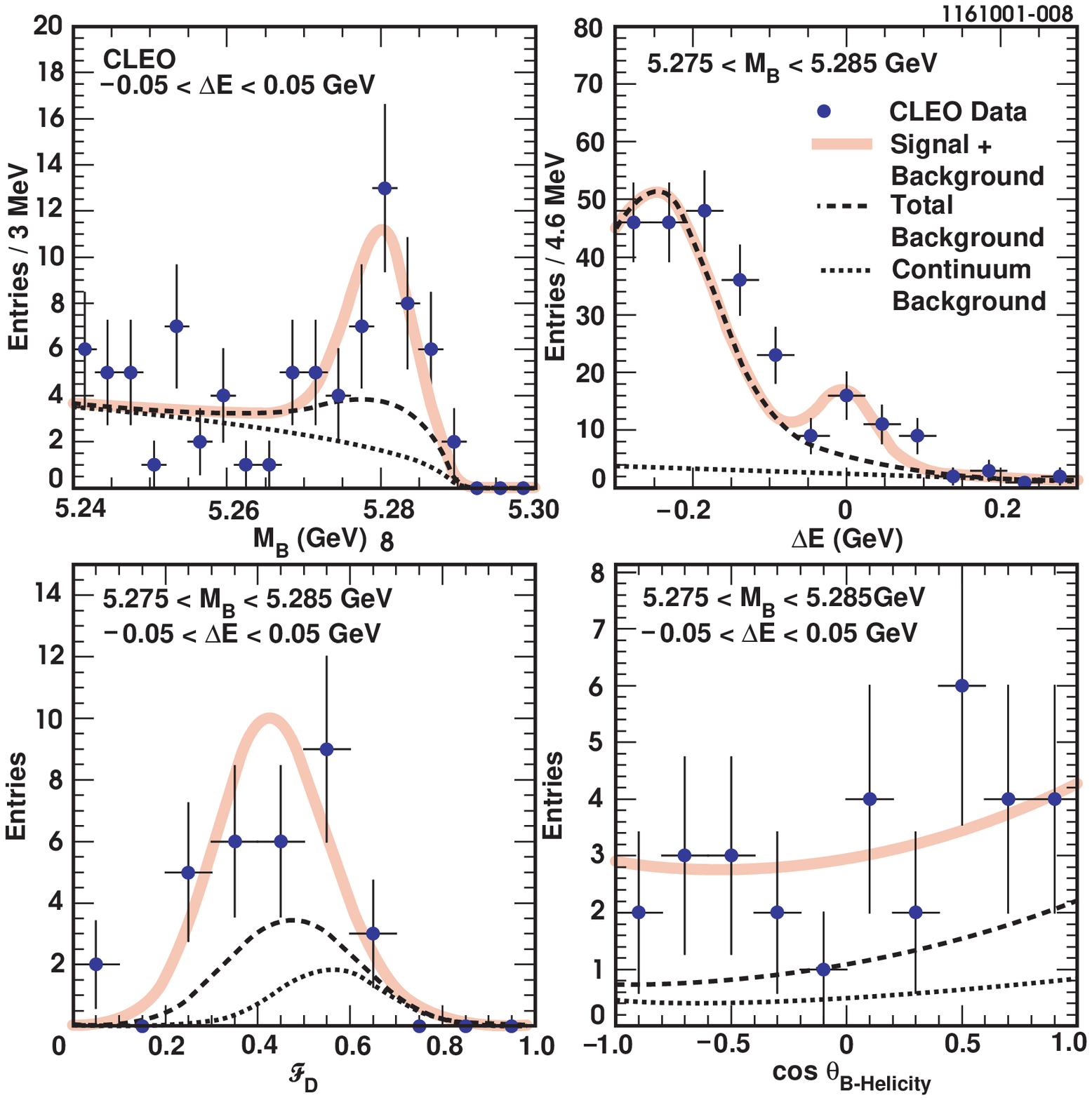,width=\myfigwidth}}
\begin{center}
\parbox{145mm}{\caption[ ]{\label{signalregion_overlay_bdstpi0.eps}\it Distribution of fit
input variables for \Btodstpi. The results of the unbinned, extended
maximum likelihood fit are shown as the 
full line. The dotted line represents the fitted 
continuum and the dashed line is the fit result for the sum of \BBar\ 
and continuum background. 
To enhance the signal for display purposes, the fit results are projected 
into the \mbc-\deltae\ signal region.}}
\end{center}
\end{figure}

We consider sources of systematic uncertainties from the PDF shapes, D
and \yfours\ branching ratios, luminosity, possible 
fit bias, B candidate reconstruction and cross-feed between different 
modes. The dominant systematic 
uncertainty comes from the PDF shapes. The systematic 
uncertainty on the shapes is 
derived by varying the PDF shapes within the statistical errors of the fit 
parametrization as well as comparing the CLEO data in 
the \deltae\ and \mbc\ sideband regions to the PDF shapes and taking differences as systematic errors.
Figures \ref{signalregion_overlay_bdpi0.eps} and 
\ref{signalregion_overlay_bdstpi0.eps} show our results 
for \Btoddstpi\ with the number of signal, \BBar\ and Continuum background as 
free parameters of the fit.
The fit result is projected into a 
signal region, defined in the \mbc-\deltae\ plane as \dsignalregion. 
The fit results describe the data well. The background in the sidebands
is also well modeled by the fit. 
\begin{table}[htb]
\begin{center}
\begin{tabular}{l|l|lcrrr}
& 			 		& Fit Yield & S{\small
ignifi-} & \Efficiency\ \ & \BR(D$^{(*)}$) & \BR(D$^{(*)0}$\gpz)\\
& \up{Mode} & (Events) & {\small cance}($\sigma$) & (\%) & (\%) & (10$^{-4}$)\\ \hline
\hline
		&\Dtokpi 		& \Res{37.5}{6.9}{7.2}	& 8.5 &\efa& 3.82  & 2.74$\pm$0.53 \\
\Btodpi 	&\Dtokpipi		& \Res{42.1}{8.6}{9.0}	& 6.8 &\efb& 12.94 & 2.49$\pm$0.53 \\
		&\Dtokthreepi		& \Res{44.6}{10.2}{10.7}& 5.3 &\efc& 7.48  & 3.25$\pm$0.78 \\
\hline
\multicolumn{2}{l|}{Averaged \BR(\Btodpi)\lamplace}    & & \Bdpisig &&&\BRdpi \\
\hline
		&\Dsttodpi, \Dtokpi	& \Res{6.8}{2.8}{3.2}	& 2.4 &\efd & 2.36  & 1.95$\pm$0.91 \\
		&\Dsttodpi, \Dtokpipi	& \Res{7.3}{3.6}{4.0}	& 2.8 &\efe
& 8.01  & 1.72$\pm$0.94 \\
\up{\Btodstpi} 	&\Dsttodpi, \Dtokthreepi& \Res{8.0}{3.7}{4.2}	& 3.1 &\eff & 4.63  & 2.21$\pm$1.15 \\
		&\Dsttodg, \Dtokpi	& \Res{6.4}{2.7}{3.0}	& 3.4 &\efg & 1.46  & 3.99$\pm$1.89 \\
\hline
\multicolumn{2}{l|}{Averaged \BR(\Btodstpi)\lamplace}	& & \Bdstpisig &   &       &  \BRdstpi   \\
\end{tabular}
\end{center}
\parbox{145mm}{\caption[ ]{\label{fityield.tab}Fit yields for all decay 
modes. Our results are based on the D branching ratios given in column
5 \cite{PDG}. 
Our measurement of the B branching ratios is given in the last column.}}
\end{table}

The observation of \Btoddstpi\ completes 
the measurement of D$^{(*)}\pi$ final states. This 
allows us 
to calculate the relative phase between the 
isospin 1/2 and 3/2 amplitudes in the D${^{(*)}\rm \pi}$ system. The basic relation can 
be expressed in an amplitude triangle: 
${\cal A}(\aDz\Pgpp)={\cal A}(\PDm\Pgpp)+\sqrt{2}\:{\cal A}(\aDz\gpz)$,
following the formulation in \cite{rosner99}.
With the PDG values \cite{PDG}
\BR(\aDz\Pgpp) = (53$\pm$5)\kreuz 10$^{-4}$, \BR(\PDm\Pgpp) = (30$\pm$4)\kreuz 10$^{-4}$, \BR(\aDstz\Pgpp) = (46$\pm$4)\kreuz 10$^{-4}$, \BR(${\rm D^{*-}}$\Pgpp) = (27.6$\pm$2.1)\kreuz 10$^{-4}$, $\tau(B^+)/\tau(B^0)$ = $1.073\pm0.027$,
and our measurement of \Btoddstpi,
we determine the relative phase between the isospin amplitudes to be
$\cos \delta_{I,D} =$\cosdi\ for the ${\rm D\pi}$ final state and
$\cos \delta_{I,D*}$=\cosdistar\ for ${\rm D^*\pi}$.
The ratios of isospin amplitudes ${\rm A_{1/2}/A_{3/2}}$ are \amplratiodpi\
(${\rm D\pi}$) and \amplratiodstpi\ (${\rm D^*\pi}$).
A similar calculation has been performed in \cite{neubert_petrov_color_suppr} using our preliminary results \cite{cleo_eps01} and preliminary results
obtained by the Belle collaboration \cite{belle_eps01}.

Models of hadronic B decay \cite{heavy_flavors} have successfully
described experimental results using two phenomenological parameters,
$a_1$ and $a_2$, that characterize non-factorizable contributions. Both are
believed to be process-dependent but so far experimental data have
been consistent with universal values for $a_1$ and $a_2$.
Recent work by Beneke, Buchalla, Neubert and Sachrajda \cite{neubert_beneke} 
has shown that $a_1$ is only slightly process-dependent.
Based on our \Btodpi\ measurement, we derive a value $a_2$=0.57$\pm$0.06.
Comparing our result to the $a_2$ value from 
two-body B decays to charmonium, $a_2$=0.29 \cite{heavy_flavors}, the
process dependence of $a_2$ is favored \cite{neubert_petrov_color_suppr}.

To summarize we observed the color-suppressed decays \Btodpi\ and \Btodstpi. 
The number of signal events in our data sample was obtained from an 
unbinned extended maximum likelihood fit in four variables.
The measurements of the two branching fractions are \BR(\Btodpi) = (\BRdpi$\pm$\syserrBdpi)\kreuz 10$^{-4}$ and \BR(\Btodstpi) = (\BRdstpi$\pm$\syserrBdstpi)\kreuz 10$^{-4}$. The first error is statistical and the second systematic.
The statistical significance of the \Dz\gpz\ signal is \Bdpisig$\sigma$
(\Bdstpisig$\sigma$ for \Dstz\gpz).

We gratefully acknowledge the effort of the CESR staff in providing us with
excellent luminosity and running conditions.
This work was supported by 
the National Science Foundation,
the U.S. Department of Energy,
the Research Corporation
and the Texas Advanced Research Program.


\begin{thebibliography}{99}
\bibitem{browder_honscheid} 
T.~E.~Browder, K.~Honscheid and D.~Pedrini,
Ann.~Rev.~Nucl.~Part.~Sci.~{\bf 46}, 395 (1996).
\bibitem{heavy_flavors} M.~Neubert and B.~Stech in {\it Heavy Flavors}, 
edited by A.J.~Buras and M.~Lindner (World Scientific, Singapore, 
2nd edition 1998). 
\bibitem{rosner99} 
J.L.~Rosner, Phys.~Rev.~D {\bf 60}, 074029 (1999).
\bibitem{bdpi0_cleo2} 
CLEO Collaboration, {B. Nemati \etal}, Phys.~Rev.~D {\bf 57}, 5363 (1998).
\bibitem{cleo2_detector} CLEO Collaboration, Y.~Kubota \etal, Nucl.~Instrum.~Methods Phys.~Res., {\bf A320}, 66 (1992).
\bibitem{cleo2.5_detector} T.~Hill, Nucl.~Instrum.~Methods Phys.~Res., Sect.~{\bf A418}, 32 (1998). 
\bibitem{cleo2_pipikk} CLEO Collaboration, Phys.~Rev.~Lett.~{\bf 85}, 515 (2000).
\bibitem{sphericity} S.L.~Wu, Phys.~Rep.~C {\bf 107}, 59 (1984). 
\bibitem{bifurcated} A bifurcated Gaussian is a Gaussian with
different widths for the left and right part of the curve.
\bibitem{fplusminuserror} CLEO Collaboration, Phys.Rev.Lett.~{\bf 86}, 2737 (2001). 
\bibitem{PDG} D.E.~Groom \etal, (Particle Data Group), Eur.~Phys.~Jour.~{\bf C15}, 1 (2000) and 2001 partial update for edition 2002 (URL: http://pdg.lbl.gov).
\bibitem{neubert_petrov_color_suppr} M.~Neubert and A.~Petrov, 
Phys.~Lett.~{\bf B519}, 50 (2001).
\bibitem{cleo_eps01} E.~von Toerne, in {\it Proceedings of International Europhysics Conference
on High Energy Physics}, Budapest 2001, (to be published).
\bibitem{belle_eps01} R-S.~Lu, in {\it Proceedings of International Europhysics Conference
on High Energy Physics}, Budapest, (to be published), and hep-ex/0107048,
Final results by the Belle Collaboration: hep-ex/0109021 (to be published).
\bibitem{neubert_beneke} 
M. Beneke \etal, Nucl.Phys.~{\bf B591} (2000) 313.
\end{thebibliography}
\end{document}